\documentclass{llncs}
\usepackage{amsfonts}
\usepackage{bm}
\newcommand{\C}[1]{{\cal {#1}}}
\newcommand{\m}[1]{{\bf  {#1}}}
\newcommand{\tr}{^{\sf T}}
\title{A Continuous Refinement Strategy for the Multilevel Computation
of Vertex Separators}
\author{William W. Hager\inst{1} \and James T. Hungerford\inst{1} 
\and Ilya Safro\inst{2}}
\institute{Department of Mathematics, University of Florida, Gainesville, FL \and School of Computing, Clemson University, Clemson, SC}
\begin{document}
\maketitle
\begin{abstract}
The Vertex Separator Problem (VSP) on a graph is the problem of finding the smallest
collection of vertices whose removal separates the graph into two disjoint
subsets of roughly equal size. Recently, Hager and Hungerford 
\cite{HagerHungerford12} developed a continuous bilinear programming formulation
of the VSP. In this paper, we reinforce the bilinear programming approach with a
multilevel scheme for learning the structure of the graph.

 
\end{abstract}

\section{Introduction}\label{sectIntro}

Let $G = (\C{V},\C{E})$ be an undirected 
graph with vertex set $\C{V}$ and edge set $\C{E}$.
Vertices are labeled $1$, $2$, $\ldots$, $n$. 
We assign to each vertex a non-negative weight $c_i\in\mathbb{R}_{\ge 0}$. If
$\C{Z}\subset\C{V}$, then we let $\C{W}(\C{Z}) = \sum_{i\in\C{Z}} c_i$ be
the total weight of vertices in $\C{Z}$.
Throughout the paper, we assume that $G$ is simple;
that is, there are no loops or multiple edges between vertices.  

The Vertex Separator Problem (VSP) on $G$ is to find the smallest weight
subset $\C{S}\subset\C{V}$ whose removal separates the graph into two roughly
equal sized subsets $\C{A}$, $\C{B}\subset\C{V}$ such
that there are no edges between $\C{A}$ and $\C{B}$; that is,
$(\C{A}\times\C{B})\cap\C{E} = \emptyset$. We may formulate the
VSP as
\begin{eqnarray}
& \displaystyle{\min_{\C{A},\C{B},\C{S}\subset\C{V}}\quad \C{W}(\C{S})}\nonumber\\
& \displaystyle{\mbox{subject to}\quad \C{A}\cap\C{B} = \emptyset,\quad
(\C{A}\times\C{B})\cap\C{E} = \emptyset}, \quad\label{VSP}\\
& \displaystyle{\quad \ell_a \le |\C{A}| \le u_a, \quad \mbox{and}\quad
\ell_b \le |\C{B}| \le u_b \enspace} \nonumber.
\end{eqnarray}
Here, the size constraints on $\C{A}$ and $\C{B}$ take the form of upper
and lower bounds. 
Since the weight of an optimal separator $\C{S}$ is typically
small, in practice 
the lower bounds on $\C{A}$ and $\C{B}$ are almost never attained at
an optimal solution, and may be taken to be quite small. 
In \cite{BalasSouza2005}, the authors consider the case
where $\ell_a = \ell_b = 1$ and $u_a = u_b = \frac{2n}{3}$ for the development
of efficient divide and conquer algorithms.
The VSP has several applications, including parallel computations 
\cite{PothenSimonLiou90},  
VLSI design \cite{Ullman84,KernighanLin70}, 
and network security. Like most graph partitioning 
problems, the VSP is NP-hard \cite{BuiJones92}.
Heuristic methods proposed
include vertex swapping algorithms \cite{KernighanLin70,FiducciaMattheyses82}, 
spectral methods \cite{PothenSimonLiou90}, continuous 
bilinear programming \cite{HagerHungerford12}, 
and semidefinite programming \cite{FeigeHajiaghayiLee2008}. 

For large-scale graphs, heuristics are more effective
when reinforced by a multilevel framework: first coarsen the graph to a 
suitably small size; then, solve the problem for the coarse graph; and finally,
uncoarsen the solution and refine it to obtain a solution for the original 
graph \cite{RonSB11}. Many different multilevel frameworks have been proposed in the past two decades \cite{bmsss13}.  
One of the most crucial parameters in a multilevel algorithm is the choice of 
the refinement scheme.
Most multilevel graph partitioners and VSP solvers refine
solutions using variants of the Kernighan-Lin \cite{KernighanLin70} or 
Fidducia-Matheyses \cite{FiducciaMattheyses82,LeisersonLewis87} algorithms. 
In these algorithms, a low weight
edge cut is found by making a series of vertex swaps starting from an
initial partition, and a vertex separator is obtained by selecting vertices 
incident to the edges in the cut.
One disadvantage of using these schemes is that they assume that
an optimal vertex separator lies near an optimal edge cut.
As pointed out in \cite{FeigeHajiaghayiLee2008}, 
this assumption need not hold in general. 

In this article, we present a new refinement strategy for multilevel
separator algorithms which computes vertex separators directly. Refinements 
are based on solving the following continuous bilinear program (CBP):
\begin{eqnarray}\label{CVSP}
& \displaystyle{\max_{\m{x},\m{y}\in\mathbb{R}^n}\quad 
\m{c}\tr(\m{x} + \m{y}) - \gamma \m{x}\tr(\m{A} + \m{I})\m{y}}\\
& \displaystyle{\mbox{subject to}\quad \m{0}\le\m{x}\le\m{1},\quad
\m{0}\le\m{y}\le\m{1},\quad\ell_a \le \m{1}\tr\m{x} \le u_a,\quad \mbox{and} \quad
\ell_b \le \m{1}\tr\m{y}\le u_b}\enspace \nonumber.
\end{eqnarray}
Here, $\m{A}$ denotes the adjacency matrix for $G$ (defined by 
$a_{ij} = 1$ if $(i,j) \in\C{E}$ and $a_{ij} = 0$ otherwise), $\m{I}$ is the
$n\times n$ identity matrix, 
$\m{c}\in\mathbb{R}^n$ stores the vertex weights, and 
$\gamma := \max\;\{c_i : i\in\C{V} \}$.
In \cite{HagerHungerford12}, the authors show that (\ref{CVSP}) is
equivalent to (\ref{VSP}) in the following sense:
Given any feasible point
$(\hat{\m{x}},\hat{\m{y}})$ of (\ref{CVSP}) one can find a piecewise linear path
to another feasible point $(\m{x},\m{y})$ such that 
\begin{equation}\label{props}
f(\m{x},\m{y})\ge f(\hat{\m{x}},\hat{\m{y}}),\quad
\m{x},\m{y}\in\{0,1\}^n, \quad \mbox{and}\quad
\m{x}\tr(\m{A} + \m{I})\m{y} = 0 \enspace .
\end{equation}
(see the proof of Theorem 2.1, \cite{HagerHungerford12}).
In particular, there exists a global solution to (\ref{CVSP}) satisfying
(\ref{props}), and for any such solution, an optimal solution to 
(\ref{VSP}) is given by 
\begin{equation}\label{partition}
\C{A} = \{ i : x_i = 1\},\quad \C{B} = \{i : y_i = 1\},\quad
\C{S} = \{i : x_i = y_i = 0\} \enspace .
\end{equation}
(Note that the fact that (\ref{partition}) is a partition of $\C{V}$ with 
$(\C{A}\times\C{B})\cap\C{E} = \emptyset$ follows from the
last property of (\ref{props}).)

In the next section, we outline a multilevel algorithm which incorporates 
(\ref{CVSP}) in the refinement phase.
Section 3 concludes the paper with some computational results 
comparing the effectiveness of this refinement 
strategy with traditional Kernighan-Lin refinements.

\section{Algorithm}
The graph $G$ is coarsened by visiting each vertex and matching \cite{bmsss13}
it with an unmatched neighbor to which it is most strongly coupled. 
The strength of the coupling between vertices is measured using a heavy edge 
distance: For the finest graph, all edges are assigned a weight equal to $1$;
as the graph is coarsened, multiple edges arising between any two vertex aggregates
are combined into a single edge which is assigned a weight equal to the
sum of the weights of the constituent edges.
This process is applied recursively: first the finest graph is coarsened,
then the coarse graph is coarsened again, and so on. 
When the graph has a suitably small size, the coarsening stops and
the VSP is solved for the coarse graph using any available method 
(the bilinear program (\ref{CVSP}), Kernighan-Lin, etc.) The solution is stored as a pair of incidence vectors
$(\m{x}^{\rm coarse},\m{y}^{\rm coarse})$ for $\C{A}$ and $\C{B}$ (see (\ref{partition})).

When the graph is uncoarsened, 
$(\m{x}^{\rm coarse},\m{y}^{\rm coarse})$ yields
a vertex separator for the next finer level by assigning components of $\m{x}^{\rm fine}$
and $\m{y}^{\rm fine}$ to be equal to $1$ whenever their counterparts in the coarse graph
were equal to $1$, and similarly for the components equal to $0$. This initial 
solution is refined by alternately holding $\m{x}$ or $\m{y}$ fixed, 
while solving (\ref{CVSP}) over the free variable and taking
a step in the direction of the solution. (Note that when $\m{x}$ or $\m{y}$ is
fixed, (\ref{CVSP}) is a linear program in the free variable, and thus can be
solved efficiently.) When no further improvement is
possible in either variable, the refinement phase terminates and a separator is
retrieved by moving to a point $(\m{x},\m{y})$ which satisfies (\ref{props}). 

Many multilevel algorithms employ techniques for escaping false local optima 
encountered during the refinement phase. For example, 
in \cite{SafroRonBrandt06b} 
simulated annealing is used. In the current algorithm, local maxima are  
escaped by reducing the penalty parameter 
$\gamma$ from its initial value of $\max\;\{c_i : i\in\C{V}\}$.
The reduced problem is solved using 
the current
solution as a starting guess. 
If the current solution is escaped, then
$\gamma$ is returned to its initial value and the refinement phase is repeated. 
Otherwise, $\gamma$ is reduced in small increments until it reaches 
$0$ and the escape phase terminates. 

\section{Computational Results}
The algorithm was implemented in C++. Graph structures such as the adjacency 
matrix and the vertex weights were stored using the LEMON Graph Library 
\cite{LEMON}.
For our preliminary experiments, we used several symmetric matrices from the 
University of Florida Sparse Matrix Library having dimensions between $1000$ and
$5000$. For all problems, we used the parameters $\ell_a = \ell_b = 1$, 
$u_a = u_b = \lfloor{0.503n}\rfloor$, and $c_i = 1$ for each 
$i = 1,\;2,\;\ldots\;, n$. 
We compared the sizes of the separators obtained by our algorithm with the 
routine METIS\underline{\hspace{0.25cm}}ComputeVertexSeparator 
available from METIS 5.1.0. 
Comparsions are given in Table \ref{table}.



\begin{table}
\begin{center}
\begin{tabular}{|l c c c c | l c c c c|}
\hline
Problem & $|\C{V}|$ & Sparsity & CBP & METIS & Problem & $|\C{V}|$ & Sparsity & CBP & METIS\\
\hline
{\sc bcspwr09 }  & 1723  & .0016 & 8    & 7  & 
{\sc G42 }       & 2000  & .0059 & 498  & 489 \\
{\sc netz4504  } & 1961  & .0013 & 17   & 20 & 
{\sc lshp3466 }  & 3466  & .0017 & 61   & 61\\
{\sc sstmodel }  & 3345  & .0017 & 26   & 23 & 
{\sc minnesota } & 2642  & .0009 & 17   & 21\\
{\sc jagmesh7 }  & 1138  & .0049 & 14   & 15 & 
{\sc yeast }     & 2361  & .0024 & 196  & 229 \\
{\sc crystm01 }  & 4875  & .0042 & 65   & 65 & 
{\sc sherman1 }  & 1000  & .0028 & 28   & 32 \\
\hline
\end{tabular}
\end{center}
\caption{Illustrative comparison between separators obtained using either
METIS or CBP (\ref{CVSP})
}
\label{table}
\end{table}

Both our algorithm and the METIS routine compute vertex separators using a 
multilevel scheme. Moreover, both 
algorithms coarsen the graph using a heavy edge 
distance. Therefore, since initial solutions obtained at the coarsest level
are typically exact, the algorithms differ primarily in how the 
the solution is refined during the uncoarsening process. 
While our algorithm refines using the CBP (\ref{CVSP}), 
METIS employs Kernighan-Lin style refinements.
In half of the problems
tested, the size of the separator obtained by our algorithm was smaller than
that of METIS. No correlation was observed between problem dimension and
the quality of the solutions obtained by either algorithm. Current preliminary implementation of our algorithm is not optimized, so the running time is not compared. (However, we note that both algorithms are of the same linear complexity.) 
Nevertheless, the results in Table \ref{table} indicate that 
the bilinear program
(\ref{CVSP}) can serve as an effective refinement tool in multilevel separator
algorithms. We compared our solvers on graphs with heavy-tailed degree distributions and the results were very similar. We found that in contrast to the balanced graph partitioning \cite{bmsss13}, the practical VSP solvers are still very far from being optimal. 
We hypothesize that the breakthrough in the results for VSP lies in the combination of KL/FM and CBP refinements reinforced by a stronger coarsening scheme that introduce correct reduction in the problem dimensionality (see some ideas related to graph partitioning in \cite{bmsss13}).

\bibliographystyle{splncs}
\bibliography{library}

\end{document}